\documentclass[preprint,nofootinbib,showpacs,aps,prc]{revtex4-1}
\usepackage{graphicx}
\usepackage{amstext}
\usepackage{amssymb}

\usepackage[usenames]{color}
\begin{document}
\title{On $m_T$ dependence of femtoscopy scales \\ for meson and baryon pairs}
\author{Yu.M. Sinyukov$^{a}$, V.M. Shapoval$^a$, V.Yu. Naboka$^a$ }

\affiliation{$(a)$ Bogolyubov Institute for Theoretical Physics, Metrolohichna str. 14b, 03680 Kiev, Ukraine}

\begin{abstract}
The $m_T$-dependencies of the femto-scales, the so-called interferometry and source radii, are  investigated within the
hydrokinetic model for different types of particle pairs --- pion-pion, kaon-kaon, proton-proton and proton-lambda, --- 
produced in Pb+Pb and $p+p$ collisions at the LHC. In particular, such property of the femto-scales momentum behavior as $m_T$-scaling is studied for the systems with (w) and without (w/o) intensive transverse flow,
and also w and w/o re-scattering at the final afterburner stage of the matter evolution. The detail spatiotemporal
description obtained within hydrokinetic model is compared with the simple analytical results for the spectra and
longitudinal interferometry radii depending on the effective temperature on the hypersurface of  maximal emission,
proper time of such emission, and intensity of transverse flow. The derivation of the corresponding analytical
formulas and discussion about a possibility for their utilization by the experimentalists for the simple femtoscopy
data analysis is the main aim of this theoretical investigation.
\end{abstract}

 \maketitle



\section{Introduction}
The spatiotemporal structures of particle emission in nucleus-nucleus, proton-(anti)pro\-ton and
proton-nucleus collisions are essentially defined by the dynamics of the collision
processes~\cite{Pratt, MakSin, Sinyuk, Hama}. Therefore the correlation interferometry
\cite{Kopylov}, that measures the femtoscopic scales by means of two- (or many-) particle correlations,
allows one to study such processes experimentally.  The corresponding femtoscopic patterns can be
presented in different forms. One of them is the $k_T$-momentum dependence of the interferometry radii
$R_i(k_T=\left|{\bf p}_{T1}+{\bf p}_{T2}\right|/2)$, that results from a 3D Gaussian fit in
$q^i=p_{1}^i-p_{2}^i$ to the two-particle  correlation function $C({\bf q}, k_T)$, defined as a ratio
of the two-particle spectrum to the product of the single-particle ones. The other one is the source
function $S({\bf r^*})$ \cite{Koonin} reflecting the dependence of the pair production on the
distance ${\bf r^*}$ between the two emitted particles in the rest frame of the pair. Both patterns
supplement each other, and a reliable model should describe/predict all the mentioned types of the
femtoscopic observables, if it contains a detailed space-time picture of the collision process.

It is important to note that the correlation function behavior depends also on the particle species.
The detailed behavior of this dependence can  be used to discriminate between different scenarios
of matter evolution and particle emission in the collision processes. For example, the
hydrodynamic picture of A+A collisions for the particular case of negligible transverse flow leads
to the same $m_T^{-1/2}$  behavior
\footnote{Here, $m_T^2=m^2+(\left|{\bf p}_{T1}+{\bf p}_{T2}\right|/2)^2=m^2+k_T^2\, $
 is the transverse mass of the particle pair.} 
 of the longitudinal radii $R_l(k_T)$ for the pairs of identical pions and
kaons, and even leads to the complete  $m_T$-scaling in the case of common freeze-out
\cite{MakSin,Sinyuk}.

In Ref. \cite{Shap} it was found that hydrokinetic model (HKM) \cite{HKM, HKM1, KS, hHKM} predicts
strong violation of such a scaling between pions and kaons at the LHC, and predicts $k_T$-scaling (for
$k_T > k_0 \approx 0.4$~GeV/$c$) instead. In this letter we analyze in detail the physical reasons for
$m_T$-scaling violation. As it turned out that the reasons are quite general, there is a hope that
the found $m_T$ peculiarities will be confirmed in the LHC experimental comparative analysis of pion
and kaon femtoscopy data. In addition we predict the $m_T$-behavior for femtoscopic scales in 
other cases, including the case of meson and baryon source function radii.

To simplify the theoretical study, we reconsider the analytical results, describing $m_T$ (or
$k_T$) behavior of the femtoscopy scales in a situation with strong transverse flows typical for
RHIC and LHC energies. We derive a simple analytical formula for such a scenario, which fits well
the complex HKM results for different hadron pairs. Since the HKM describes simultaneously the large number of bulk observables, it gives experimentalists a simple
tool for estimation of the life-time of expanding fireball, related to the maximal emission of the specific hadron pair created in proton or nuclear collisions.

\section{Analytical model for the interferometry radii}

The hydrokinetic model (HKM) (see details in Ref. \cite{hHKM}) was developed to simulate the
evolution of matter formed in  relativistic heavy-ion collisions and describe/predict bulk
observables at RHIC and LHC. In Ref. \cite{hHKM} a good description was reached for pion, kaon,
(anti)proton and all charged particle spectra at different centralities, as well as for the elliptic
flows. Concerning the femtoscopic scales, the confirmed HKM prediction \cite{karpsin} about a reduction
of $\frac{R_{out}}{R_{side}}$ ratio in A+A collisions at LHC as compared to RHIC  was done;  the
interferometry radii for pions and kaons at RHIC \cite{hHKM, KS} and LHC \cite{hHKM,Shap} in A+A
collisions were calculated; the source functions for kaons and pions at RHIC were described at the
top RHIC energy and predicted for the LHC one \cite{ShapSin}. In addition, HKM describes well pion
interferometry radii in $p+p$ collisions at the LHC energy $\sqrt{s} = 7$~TeV if one incorporates the
quantum uncertainty principle into a quasi-classical event generator~\cite{PBM-Sin}.

All the results for A+A collisions are obtained at the same initial conditions at $\tau = 0.1$~fm/$c$,
produced by MC Glauber generator GLISSANDO \cite{glissando} (with just re-scaled maximal initial
energy density at the transition from RHIC to LHC), the same type of hydrodynamics evolution
(however with different  baryochemical potentials at RHIC and LHC), the same lowest temperature  when the local thermal and chemical equilibrium still exist in the hydrodynamically expanding
medium, $T_{ch}=165$~MeV, and with UrQMD cascade at the final afterburner stage. Such a self-consistent and reliable
simulation of the data, including femtoscopy ones, indicates that the spatiotemporal picture of
particle emission in the model can serve as a reference point for further studies,
no matter how much advanced
future models will be developed. Note, that in HKM and similar models many factors act
simultaneously, the results can be obtained only by means of time-consuming numerical calculations, and
it is not always clear, how each physical factor affects the complete space-time picture of the
particle emission in the collision process.

Here we provide some analytical estimates for the spectra and femtoscopy scales, aiming to reveal the main
parameters forming the particle emission distribution in space and time. In Refs.~\cite{HKM1,CF} the
problem of spectra formation in hydrodynamic approach to A+A
collisions is considered within the Boltzmann equations. It is shown analytically and illustrated by
numerical calculations that the particle momentum spectra can be described by the Cooper-Frye
prescription (CFp) \cite{Cooper} despite freeze-out is not
sharp and has the finite temporal width. The latter is equal to the inverse of the particle
collision rate at 4-points ($t_{\sigma}({\bf r}, p), {\bf r})$ of the maximal particle emission at a
fixed momentum $p$. The set of these points forms the hypersurfaces
$t_{\sigma}({\bf r},p$) which depend on the values of $p$ and typically do not enclose
completely the initially dense matter. This is an important difference
from the standard CFp, with a common freeze-out
hypersurface (that is, typically, an isotherm) for all the momenta $p$.
Also, the well known problem of CFp -- negative contributions to the
spectra from non-space-like parts of the freeze-out hypersurface -- is naturally eliminated in this
improved/generalized prescription: at each momentum $p$ the hypersurface $\sigma_p$ is always a
space-like one.

In this letter we partially use the results of Refs.~\cite{AkkSin1,AkkSin2} and consider the hypersurface of
constant Bjorken (proper) time $\tau$, $\sigma_p$: $\tau = const$, as the hypersurface of maximal
emission for soft enough particle momenta $p$. It is shown \cite{CF} that such a hypersurface,
limited in transverse direction ${\bf r}_T$, is typical for soft particles. For the particles with hard
momenta, e.~g. $p_T > 0.8$~GeV/$c$,
the hypersurface of maximal emission is different from $\tau \approx const$: there is the correlation
between the transverse radius and the time of radiation (see details in \cite{AkkSin1,AkkSin2}). To cut the hypersurface of
maximal emission (m.~e.) $\tau = \tau_{m.~e.}=const$ in the transverse plane for soft quanta, and at the
same time get analytical approximation, we use here the Gaussian cutoff factor $\rho({\bf r}_T)$
in the way proposed first in Refs. \cite{AkkSin1,AkkSin2}. Namely, the Wigner function for bosons in almost central events has the
form:
\begin{equation}
f_{l.eq.}(x,p)=\frac{1}{(2\pi)^3}\left[\exp(\beta p\cdot u(\tau_{m.e.},{\bf r}_T) -\beta\mu)-1\right]^{-1}\rho({\bf r}_T).
\label{wigner}
\end{equation}
Here and below $\beta = 1/T$ is the inverse of temperature at $\tau_{m.e.}=\sqrt{t^2_{m.e.}-x^2_{L, m.e.}}$,
${\bf r}_T\equiv {\bf x}_T$ is 2D transverse radius-vector, 4-coordinate is $x^{\mu}=(\tau\cosh\eta_L,{\bf
r}_T,\tau\sinh\eta_L)$ with longitudinal rapidity $\eta_L= \text{arctanh}\,v_L=\frac{1}{2}\ln\frac{t+x_L}{t-x_L}$,
transverse rapidity $\eta_T \equiv \eta_T(r_T) = \text{arctanh}\,v_T(r_T)$, the
hydrodynamic
velocity $u^{\mu}(x)=(\cosh\eta_L\cosh\eta_T,\frac{{\bf r}_T}{r_T}\sinh\eta_T,
\sinh\eta_L\cosh\eta_T)$. The cutoff factor $\rho({\bf r}_T)$ has the form~\cite{AkkSin1,AkkSin2}:
\begin{equation}
\rho({\bf r}_T)=\exp[-\alpha (\cosh\eta_T(r_T)-1)].
\label{rho}
\end{equation}
Here, according to results of \cite{AkkSin1,AkkSin2} and the interpretation corrected for the improved CFp~\cite{CF}, the parameter $\alpha = R_v^2/R_T^2$, where $R_T$ is the transverse homogeneity length
in $r_T$ (near $r_T=0$ for small $k_T$ ) along the hypersurface $\tau = \tau_{m.e.}=const$, 
and $R_v$ is the hydrodynamic
length, $R_v= (v^{\prime}(r_T))^{-1}$, near the same $r_T$. The small $\alpha$ can be reached at 
very intensive flow when the hydrodynamic length $R_v$  is much smaller than the homogeneity length 
$R_T$ (which is, generally speaking, different for different particles), while the upper limit is reached 
in the case of no transverse flow, $R_v=\infty$. One can see that the contributions from particles with high momenta emitted from the
hypersurface $\tau = \tau_{m.e.}=const$ are suppressed by the cutoff factor, since such particles
radiate typically from rapidly moving fluid elements, $\cosh\eta_T(r_T) \gg 1$. Another $\tau({\bf r}_T)$ hypersurface should be used for hard particles.

To get single particle spectra $p_{0}d^{3}N/d^{3}p$
and the correlation function of bosons  $C(p,q)$ in smoothness \& mass shell approximation, we use the
generalized Cooper-Frye method as it was discussed above:
\begin{eqnarray}
p_{0}\frac{d^{3}N}{d^{3}p}=\int_{\sigma_{m.e.}(p)}d\sigma_{\mu} p^{\mu} f_{l.eq.}(x,p),
\label{sp-def}
\end{eqnarray}

\begin{equation}
C(p,q)\approx 1+\frac{\left|\int_{\sigma_{m.e.}(k)} d\sigma_{\mu}k^{\mu} f_{l.eq.}(x,k)\exp(iqx)\right|^{2}}{\left(\int_{\sigma_{m.e.}(k)}
d\sigma_{\mu}k^{\mu} f_{l.eq.}(x,k)\right)^{2}},
\label{corr-approx}
\end{equation}
where
$q=p_{1}-p_{2}$, $k^{\mu}=(\sqrt{m^2+\left(\frac{\mathbf{p_{1}}+\mathbf{p_{2}}}{2}\right)^2},\frac{\mathbf{p_{1}}+\mathbf{p_{2}}}{2})$.
Note that in smoothness \& mass shell approximation $k\approx p=(p_1+p_2)/2$. Then 4-vector $q$ has
three independent components, which can be selected along the beam axis, $q_{long}\equiv q_l$, along
the pair transverse momentum vector ${\bf k}_T$, $q_{out}\equiv q_o$, and along {\it side} direction, $q_{side}\equiv q_s$, being orthogonal to both {\it long} and {\it out} directions.

With all these notations, one can apply the saddle point method (in complex plane) to calculate
the spectra and correlation function (\ref{corr-approx}) in Boltzmann approximation for longitudinally boost-invariant expansion. The results were obtained in Ref. \cite{Tolstykh} on the base of the approach
using the Wigner function (\ref{wigner}) \cite{AkkSin1,AkkSin2}. The remarkable fact is that the behavior of
the correlation function in {\it long} direction depends solely on the parameter $\alpha$ and 
does not depend on the transverse velocity profile at the 
hypersurface of maximal emission, while, as expected, the behavior of
the correlation function in {\it out} and {\it side} directions is model dependent and quite
sensitive to the details of the velocity profile. Because of this, here we analyze and use only the
longitudinal projection of the correlation function. Let us introduce the value $\lambda$, associated with the homogeneity length in longitudinal direction in a presence of transverse flow  $\lambda_l=\tau\sqrt{\frac{T}{m_T}(1-\bar{v}^2_T)^{1/2}}$ \cite{AkkSin1,AkkSin2}:
\begin{equation}
\lambda^2 =\frac{\lambda_l^2}{\tau^2}=\frac{T}{m_T}(1-\bar{v}^2_T)^{1/2},
\label{lambda}
\end{equation}
where the transverse velocity in the saddle point $\bar{v}_T=k_T/(m_T+\alpha T)$. Here $T=T_{m.e.}$
is the temperature at the hypersurface of maximal emission, $\tau=\tau_{m.e.}$.
Then in LCMS \cite{Tolstykh}:
\begin{equation}
C(k,q_l,q_s=q_o=0)=1+\frac{\exp\left[\frac{2}{\lambda^2}\left(1-\sqrt{1+\tau^2\lambda^4q_l^2}\right)\right]}{\left[1+\tau^2\lambda^4q_l^2\right]^{3/2}}
\stackrel{k_T\rightarrow \infty}{\longrightarrow} 1+\exp(-\lambda_l^2 q_l^2).
\label{correlator}
\end{equation}
At large $m_T/T \gg 1$ the correlation function has the Gaussian form, the interferometry radii
coincide with homogeneity lengths  in this approximation, that was first obtained for pure Bjorken
expansion (without transverse flow: $\bar{v}_T=0$) in Ref.~\cite{MakSin}:
$R_l=\lambda_l=\tau\sqrt{\frac{T}{m_T}}$. In Ref.~\cite{Bertsch} for the same case (no transverse
flow) there was found the correction to the interferometry radii at small ratio $m_T/T \approx 1$, that
can be actual for pions:
\begin{equation}
R_l^2=\lambda_l^2 \rightarrow \lambda_l^2\frac{K_2(\frac{m_T}{T})}{K_1(\frac{m_T}{T})},
\label{Bertsch}
\end{equation}
where $K_n$ are modified Bessel functions. As one can see from (\ref{correlator}), at not large $m_T/T\approx 1$
the correlation function has non-Gaussian form in $q_l$.  Correspondingly, at not large $m_T/T$ such a radius describes only
the peak of the correlation function, where  the Gaussian approximation in the limit of small
$q_l$, $C(q_l)=1+\exp({-R_l^2q_l^2)}$, can be obtained. The detail analysis of different analytical
approximations for interferometry radii is done in Ref. \cite{Tolstykh}.

The approximation of the
correlation function (\ref{correlator}) at small $q_l$ leads to the following analytical result for the  interferometry radii in the case of boost-invariant longitudinal expansion and transverse flow with  arbitrary velocity profile:
\begin{equation}
R^2_l(k_T)=\tau^2\lambda^2\left(1+\frac{3}{2}\lambda^2\right).
\label{main}
\end{equation}
The $\lambda$ is defined by the formula (\ref{lambda}). The direct comparison with the formula
(\ref{Bertsch})  for pure Bjorken picture (in our case it corresponds to $\alpha \rightarrow \infty$)
demonstrates only $1-3\% $ deviation between the two results in all actual $k_T$ interval for pions. An
advantage of the formula~(\ref{main}) is that it is derived for the case of transverse flow of any
intensity that is especially important for LHC energies.

\section{Analytic fitting of HKM results}

\begin{figure}[!hbt]
\includegraphics[width=0.9\textwidth]{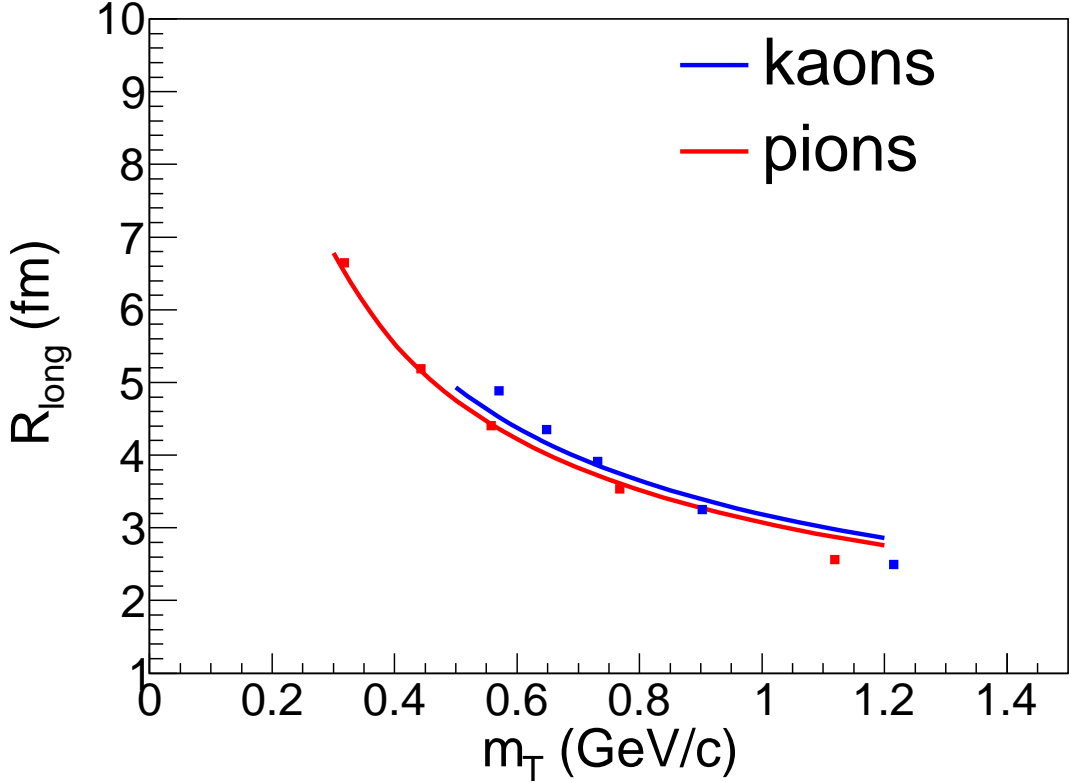}
\caption{The $m_T$ dependencies of longitudinal femtoscopy radius $R_{l}$ calculated in HKM
model without re-scattering stage for
$K^{ch}K^{ch}$ and $\pi^{-}\pi^{-}$ pairs (markers) together with the corresponding fits
(lines) according to formula~(\ref{main}) where transverse flow is absent ($\alpha=\infty$).
The temperature is $T=165$~MeV.
The results are related to $\sqrt{s_{NN}}=2.76$~GeV Pb+Pb collisions at the LHC, $c=0-5\%$, $|\eta|<0.8$, $0.14<p_T<1.5$~GeV/$c$.}
\label{mtnf}
\end{figure}

The idea of  $m_T$-scaling for the interferometry radii of different bosons has been launched  by the
1D hydrodynamic result \cite{MakSin}: $R_l\propto \sqrt{T/m_T}$.
The formulas (\ref{Bertsch}) and (\ref{main}) extended for all $m_T$ values
in the case of 1D expansion (it means $\alpha=\infty\Rightarrow \bar{v}_T = 0$
in f-la (\ref{main})) also lead to scaling: $R_l$ is the function of $\tau$ and $m_T/T$ only. In
Fig.~\ref{mtnf} we demonstrate the corresponding fit to the HKM results for pions and kaons 
with the standard Cooper-Frye prescription at the ``freeze-out'' hypersurface $T=165$~MeV, where we artificially switched off the UrQMD re-scattering, leaving only resonance
decays. As we see, the best fits are very bad: $\chi^2/\text{ndf} = 2562.31$ for pions and
$\chi^2/\text{ndf} = 585.24$ for kaons, and there is no $m_T$-scaling. The absence of $m_T$ scaling
in full HKM (with re-scattering at the afterburner UrQMD stage) for $R_l$ at LHC energies is noticed
recently in Ref.~\cite{Shap}.

\begin{figure}[!hbt]
\includegraphics[width=0.9\textwidth]{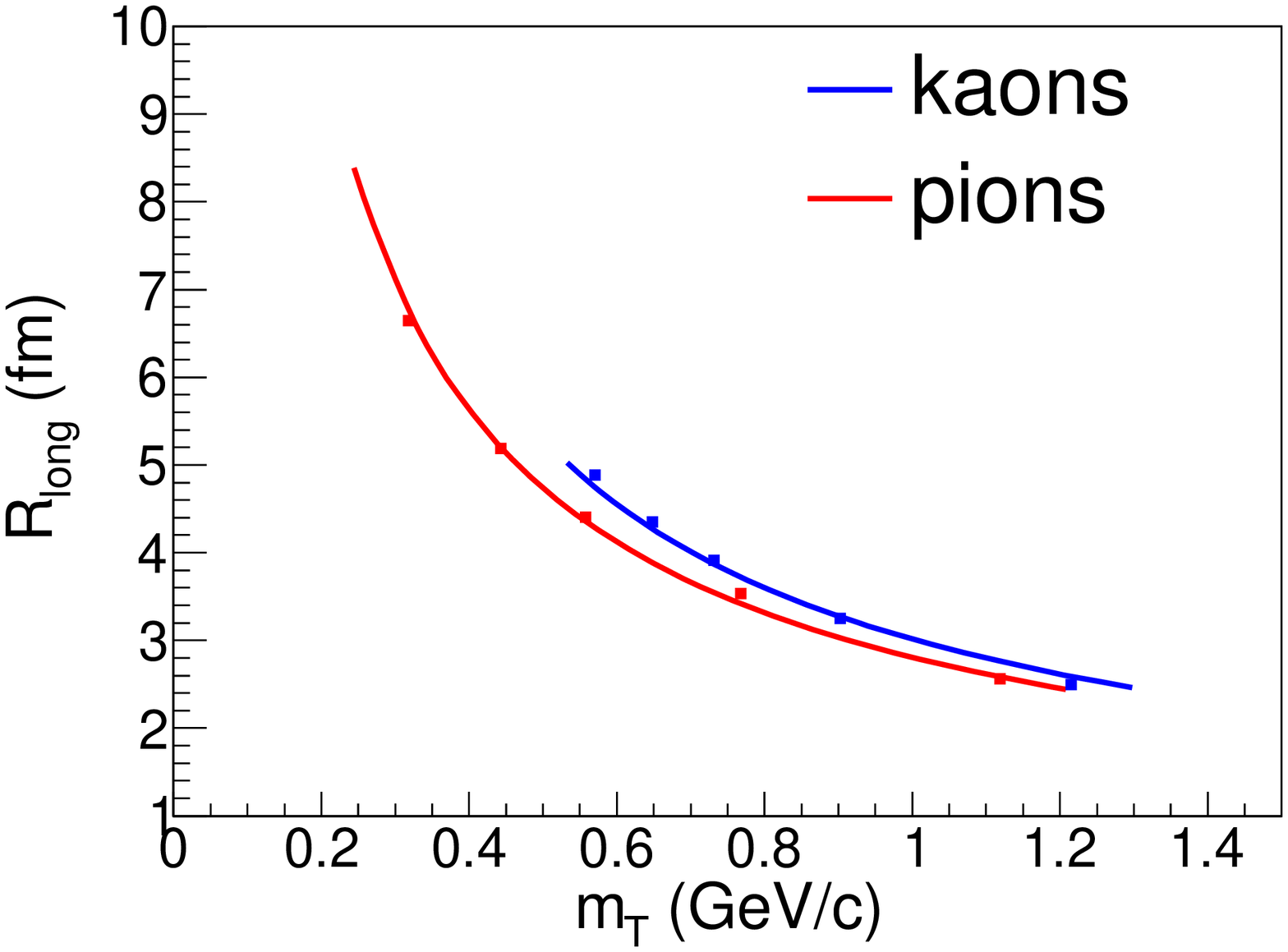}
\caption{The same as in Fig.~\ref{mtnf} but fits account now for the intensive transverse
flow, $\alpha = 2.8$.
The time parameters for pions and kaons are $\tau_{\pi}=7.41$~fm/$c$ and $\tau_{K}=7.56$~fm/$c$
correspondingly, $T_{\pi} = T_{K} = 165$~MeV.}
\label{mt1}
\end{figure}

One of the reasons for violation of scaling in HKM becomes clear from Fig.~\ref{mt1} where the same
results of ``succise'' HKM (without re-scattering) were fitted by the full f-la~(\ref{main}) that
accounts for transverse flow.
Here we also use the ``freeze-out'' temperature 165~MeV.
The fits for both pions and kaons are quite satisfactory with the same value of temperature
parameter and times
$\tau_{\pi}=7.41 $~fm/$c$ for pions and  $\tau_{K}=7.56$~fm/$c$ for kaons.
For both mesons $\alpha = 2.8$.
These fits illustrate that, as one can see from~(\ref{main}), (\ref{lambda}), the transverse
flow destroys $m_T$ scaling, since the interferometry radius depends now on both variables $m_T/T$ and $k_T/T$.

The additional reason for $m_T$-scaling violation could be re-scattering at the afterburner stage
that also can break the similarity between pion and kaon longitudinal interferometry radii. A non-Gaussian shape of the correlation functions also affects the $m_T$-behavior of
the corresponding femtoscopy scale.
To investigate in detail the role of these factors, we should analyze the results of full HKM
calculation (including the re-scattering stage).
In order to define the effective $T$ and $\alpha$ parameter values
for an analysis of the interferometry radii behavior in full HKM, one can study the
corresponding momentum spectra. These  parameters can be extracted from the combined
fit to pion and kaon spectra calculated in HKM, accounting for transverse flow.
Such a fit can be performed using the formula obtained in the same approximation as the result (\ref{correlator})~\cite{AkkSin2}:
\begin{equation}
p_0 \frac{d^3N}{d^3p} \propto \exp{[-(m_T/T + \alpha)(1-\bar{v}^2_T)^{1/2}]}.
\label{specfit}
\end{equation}

\begin{figure}[phbt]
\includegraphics[width=0.9\textwidth]{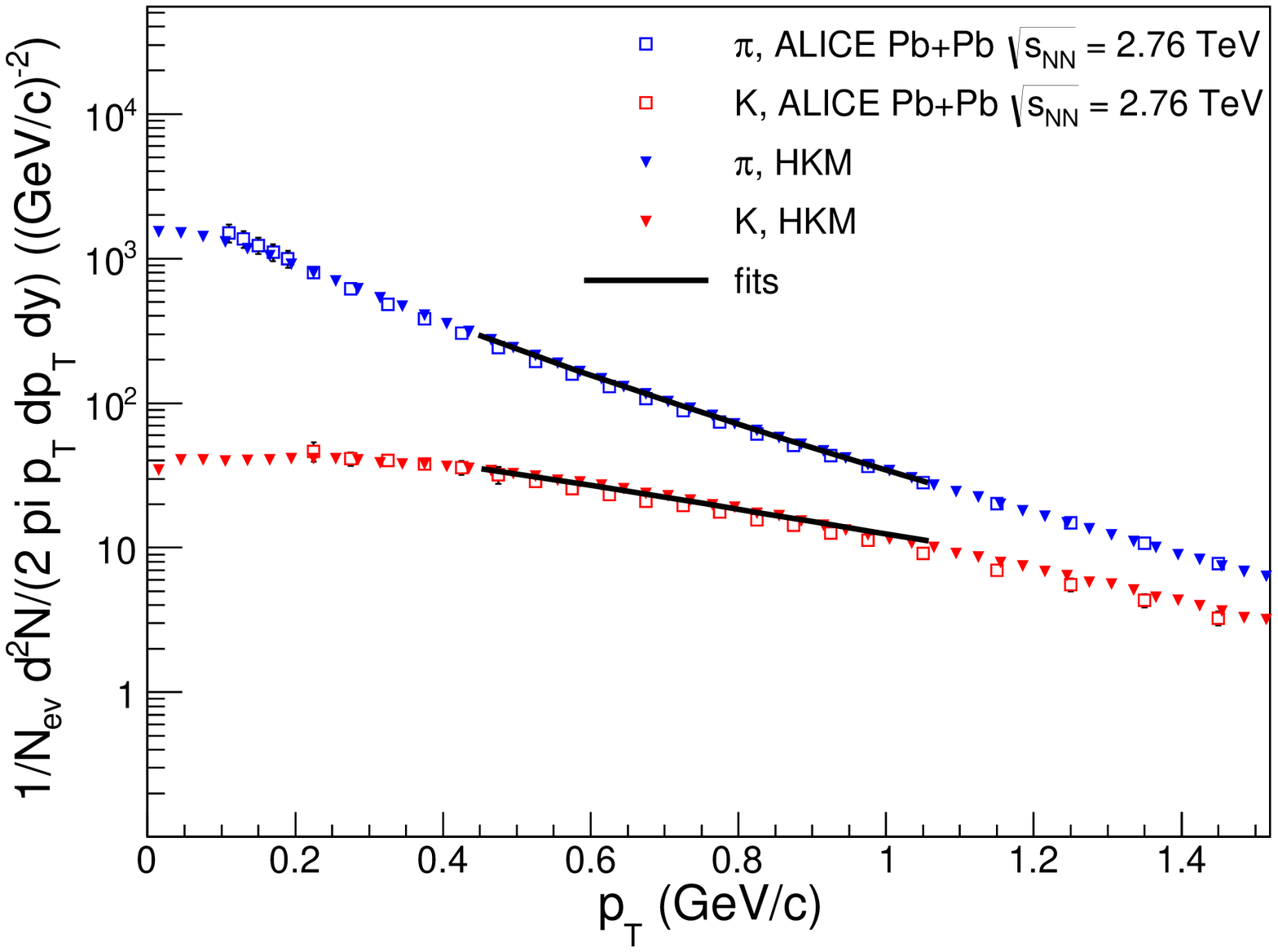}
\caption{Pion (blue) and kaon (red) momentum spectra in Pb+Pb collisions at the LHC, $\sqrt{s_{NN}}=2.76$~GeV. Open squares show experimental values \cite{AliceSpec}, triangles show HKM results,
black lines correspond to combined (with the same temperature $T$) pion and kaon spectra fit according to (\ref{specfit}).}
\label{spec}
\end{figure}

\begin{figure}[phbt]
\includegraphics[width=1.0\textwidth]{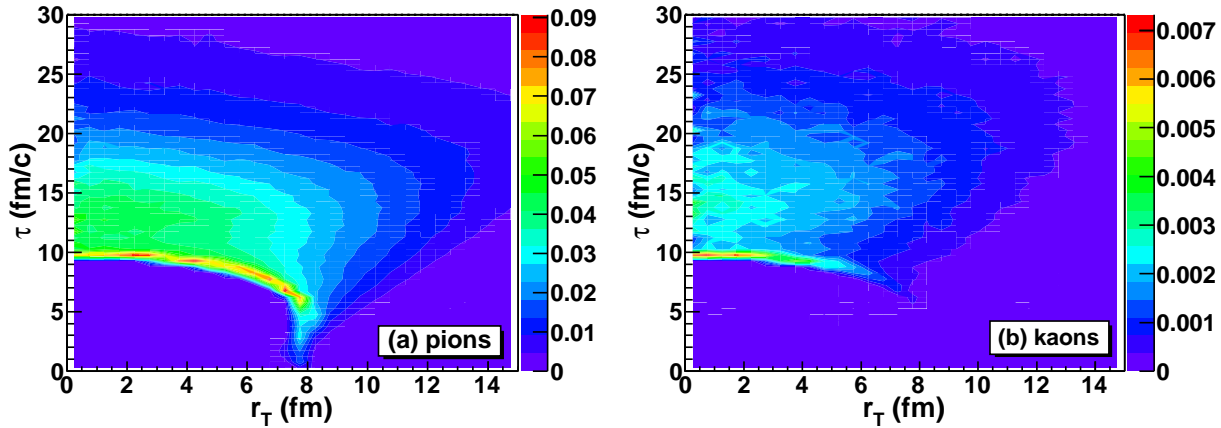}
\caption{The momentum angle averaged emission functions per units of
space-time and momentum rapidities $ g(\tau,
r_T,p_T)$ [fm$^{-3}$] (see body text) for pions~(a) and kaons~(b) obtained from
the HKM simulations of Pb+Pb collisions at the LHC
$\sqrt{s_{NN}}=2.76$~GeV, $0.2<p_T<0.3$~GeV/$c$, $|y|<0.5$,
$c=0-5$\%. } \label{emiss}
\end{figure}

\begin{figure}[phbt]
\includegraphics[width=0.9\textwidth]{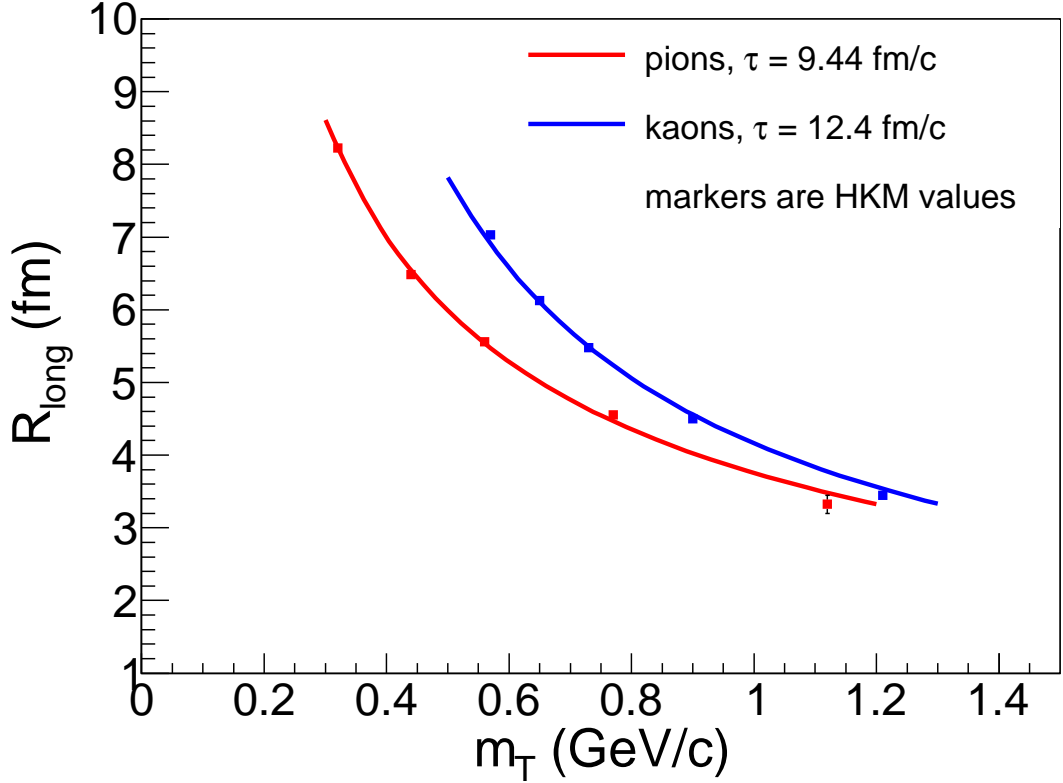}
\caption{The same as in Fig.~\ref{mtnf}, \ref{mt1} but radii are calculated
in the full HKM model including re-scattering stage. To reduce the effect of
the non-Gaussian correlation functions, we take more narrow fitting range for them,  $q=0-0.04$~GeV/$c$.
The fit parameters $T$ and $\alpha$ correspond to combined pion and kaon spectra fitting.
At $T_\pi=147$~MeV, $T_K=141$~MeV, $\alpha_\pi=8.5$ and $\alpha_K=1.5$ extracted
maximal emission times are $\tau_\pi=9.44 \pm 0.02$~fm/$c$ and $\tau_K=12.40\pm 0.04$~fm/$c$.}
\label{mttop}
\end{figure}

In Fig.~\ref{spec} one can see the experimental \cite{AliceSpec} and HKM spectra
together with the combined pion-kaon fit to HKM points. The fit with
a common temperature is performed according to (\ref{specfit}) in
$p_T$-range $0.5-1.0$~GeV/$c$. The extracted parameter values are
the following: $T=144 \pm 3$~MeV, $\alpha_\pi=5.0 \pm 3.5$ and
$\alpha_K=2.2 \pm 0.7$.
To maximally reduce the effect of the non-Gaussian correlation functions,
we restrict the fitting range to $q=0-0.04$~GeV/$c$ when extracting the interferometry radii from full HKM.
One can see the $m_T$-dependence of the radii extracted in such a way
in Fig.~\ref{mttop}
together with the analytical fit according to Eq.~(\ref{main}), where $T$, $\alpha$ are
constrained according to the combined spectra fit. The maximal emission times extracted from the fit
for pions and kaons are $\tau_\pi=9.44 \pm 0.02$~fm/$c$ and $\tau_K=12.40 \pm 0.04$~fm/$c$ respectively while
the rest of parameters take values $T_\pi=147$~MeV, $T_K=141$~MeV, $\alpha_\pi=8.5$ and $\alpha_K=1.5$.
For comparison, let us present the  pion and kaon emission pictures in HKM 
 based on the invariant emission functions $G(x,p)=p^0\frac{d^7N}{d^4x d^3p}$. In Fig.~\ref{emiss},
 we demonstrate the  reduced emission function averaged over all momentum angles
 $g(r_T,\tau; p_T)=\left.\frac{p^0 d^6N}{dr_o dr_s d\eta d\tau d p_T dy}\right|_{r_{s}=0}$ for pions and kaons
 at $0.2 < p_T < 0.3$ Gev/$c$ in the central rapidity bins for both space-time $\eta$ and momentum $y$ rapidities.
 Here $r_o$ is the component of ${\bf r}_T$ along the vector ${\bf p}_T$, and $r_s$ is its component in the
 transverse direction orthogonal to ${\bf p}_T$. Such a type of the emission function (in $r_s$ and $r_o$ variables)
  was used in \cite{AkkSin1, AkkSin2, Tolstykh} to derive the analytical approximation for the correlation function
  (see Section 2). The fit results for the times of maximal emission are in good agreement with the HKM emission
  picture. One can notice that the kaon radiation has two maxima: one happens near $\tau=10$~fm/$c$ and the
  second, broader and less pronounced, is at $\tau \approx 14-15$~fm/$c$. This second local maximum must be due to decays of $K^*(892)$, having life-time $4-5$~fm/$c$, into kaons $K^{\pm}(493.7)$.  It leads to some kind of
  mean
  time for kaons about 12 fm/$c$. It is worthy noting that the difference between this ``mean'' time and time of pion maximal emission is significantly larger than in the case when only resonance decays are taken into account, but not re-scattering. The reason is that a free streaming of fast $K^*$ with subsequent decay into $K^{\pm}$ leads to the additional large source, which contribution to the correlation function  results mostly in non-Gaussian behavior of it. At the same time the re-scattering of $K^*$ involves these mesons in some kind of collective motion producing thereby the direct link between the time of maximal emission and the value of the longitudinal radii. One can also note that the results 
  of fitting the femto-scales extracted from full HKM (accounting for the re-scattering and $K^*$ decays) as well as combined fitting of the HKM spectra give different $T$ and $\alpha$ parameters for pions and kaons, and so one can conclude that in addition to intensive transverse flows $m_T$ scaling is violated because of the re-scattering and resonance decays.

\begin{figure}[phbt]
\includegraphics[width=0.95\textwidth]{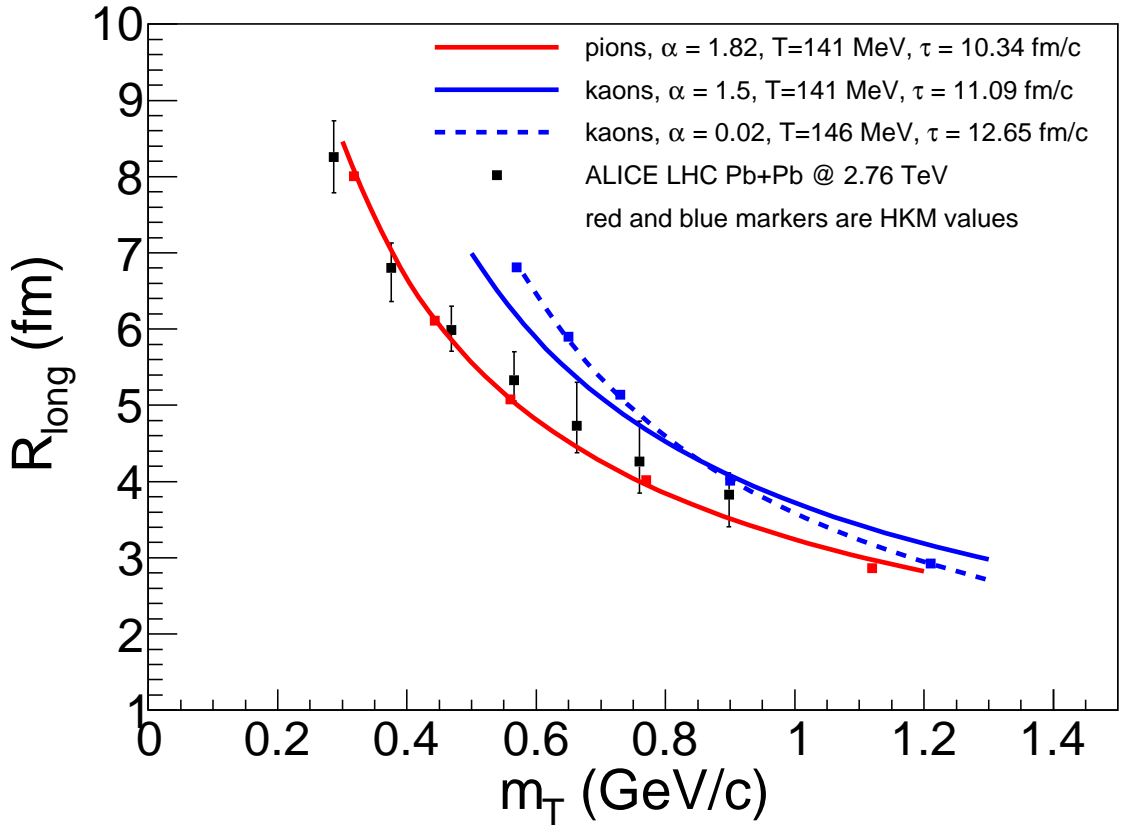}
\caption{The same as in Fig.~\ref{mttop} but the radii are
extracted from the fits to full HKM correlation functions
in a wide $q$ range, $q=0-0.2$~GeV/$c$. Thus, the significant
deviation of the CF shape from the Gaussian one leads to
a distorted radii $m_T$-behavior. At fitting it can be compensated
by decreasing $\alpha_K$ as compared to that obtained from
the spectra fit. Experimental data \cite{AliceNew} for pions are demonstrated
for comparison.}
\label{mtfull}
\end{figure}

Next, if we apply Eq.~(\ref{main}) to description of the femtoscopy
radii obtained from the fits to pion and kaon HKM correlation functions
in a broader range (e.~g.~$q=0-0.2$~GeV/$c$) \cite{Shap}, covering not only their
bell-like parts, but also the flatter ones, we will get
unsatisfactory model data description for kaons using the fit
parameters corresponding to spectra combined fit (see
Fig.~\ref{mtfull}, blue solid line) with $T_K=141$~MeV,
$\alpha_K=1.5$, and $\tau_K=11.09 \pm 0.02$~fm/$c$. As for 
pions, they are described well with parameters restricted by combined spectra fit: $T_{\pi}=141$~MeV,
$\alpha_{\pi}=1.82$, $\tau_{\pi}=10.34 \pm 0.06$~fm/$c$. To get more
or less adequate fit for the kaon longitudinal radius corresponding to wide $q$-region,
one should remove constraints on $\alpha_K$ parameter. The
dashed blue line in Fig.~\ref{mtfull} represents such a fit with
still constrained temperature, $T=144 \pm 3$~MeV, and free $\alpha_K$ and
$\tau_K$. The resulting fit at $T=146$~MeV and $\alpha=0.02$ provides good
data description and gives $\tau_K=12.65 \pm 1.58$~fm/$c$, that is
in good agreement with the previous results for radii extracted at
$q=0-0.04$~GeV/$c$ and the emission picture in Fig.~\ref{emiss}.
Therefore, the combined fit (with common temperature) for the pion
and kaon spectra leads to the reliable estimate of the maximal
particle emission times in the case of fitting only the peak of the
non-Gaussian correlation function. The same good estimates can be
extracted also from the Gaussian fit in large $q$-diapason, but for
kaons $\alpha_K$ parameter has to be fitted without restrictions
coming from common pion and kaon spectra fitting.

The difference as compared to the previous results for ``without
re-scattering'' regime as well as an essential difference between
the main fit parameters for pions and kaons mean that the
re-scattering plays an important role in violation of $m_T$ scaling,
in addition to transverse flow. The very small $\alpha$ value in
kaon fit can indicate that the re-scattering stage affects seriously
kaon emission picture and the re-scattering contribution to the femtoscopy
scale cannot be described basing on the pure hydrodynamic (plus the
resonance decays near freeze-out hypersurface) approach. The
peculiarities of the re-scattering for kaons leads to more
pronounced non-Gaussian character of the correlation function
shape than for pions and also to a distortion of the femtoscopic
scales $m_T$-dependence. It is found that for compensation of the
influence of such a factor on the extracted effective time of maximal emission, 
in the case of wide $q$-fitting range for kaons, one can eliminate restriction 
for the parameter $\alpha$ coming from combined pion and kaon spectra fit.

\begin{figure}[phbt]
\center
\includegraphics[width=0.9\textwidth]{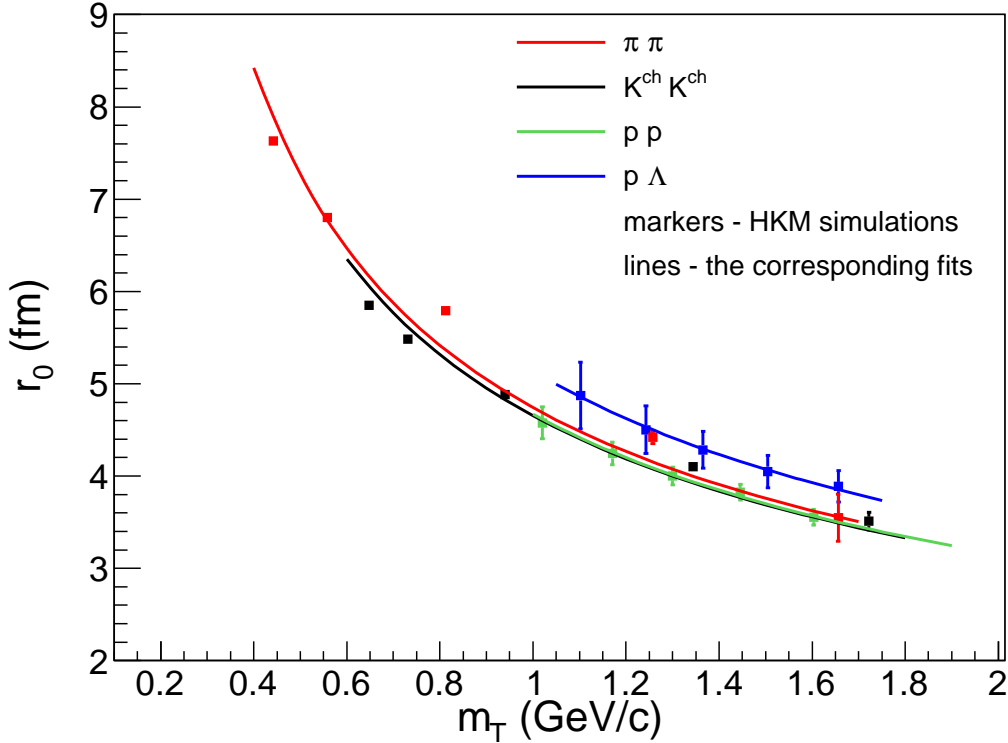}
\caption{The $m_T$ dependencies of $\pi\pi$, $K^{ch}K^{ch}$, $pp$ and $p\Lambda$
source radii $r_{0}$ extracted from corresponding angle-averaged source functions
calculated in HKM for $\sqrt{s_{NN}}=2.76$~GeV Pb+Pb collisions at the LHC, $c=0-5\%$, $|\eta|<0.8$.
Transverse momentum ranges are $0.14<p_T<1.5$~GeV/$c$ for pions and kaons, $0.7<p_T<4.0$~GeV/$c$ for protons and
$0.7<p_T<5.0$~GeV/$c$ for Lambdas.}
\label{lhcr0}
\end{figure}

In some sense a complementary method of obtaining the information about the space-time structure of
the system from
the correlation measurements, is known as the source imaging~\cite{BrownDan1,Stab,BrownDan2} aiming
to extract  the source  function $S({\bf r}^*)$, that
represents the time-integrated   distribution of particle emission points separation ${\bf r}^*$ in
the pair rest frame.
The analysis of the experimental data for baryon correlations, often based on the
Lednicky-Lyuboshitz formula \cite{LL}, supposes isotropic Gaussian
separation distribution described by single width parameter $r_0$.
This value can be extracted from the Gaussian fit
to the angle-averaged source function
$S(r)=1/(4\pi)\int_0^{2\pi} \int_0^{\pi} S(r,\theta,\phi) \sin\theta d\theta d\phi$.
In this paper we also try to apply the fitting function (\ref{main}) to pion, kaon, proton
and proton-$\Lambda$ pairs to describe the $m_T$ dependence of
source radii $r_{0}$ extracted from the corresponding angle-averaged source functions $S({\bf r}^*)$ in
 HKM by means of Gaussian fits to them, see details in Ref.~\cite{ShapSin}.
The results are presented in Fig.~\ref{lhcr0}.
At this fitting the temperature $T$ was constrained according to the
combined pion and kaon spectra fit result $T=144 \pm 3$~MeV, while
$\alpha$ and $\tau$ were left free. The fits are unexpectedly good
with the parameters $T=141$~MeV for all the particle pairs,
$\tau_{\pi\pi}=11.47 \pm 0.03$~fm/$c$, $\tau_{KK}=11.26 \pm
0.04$~fm/$c$, $\tau_{pp}=11.30 \pm 0.13$~fm/$c$,
$\tau_{p\Lambda}=12.44 \pm 0.29$~fm/$c$. The parameter $\alpha$ for
all the pairs is about $10^4$ and more, i.~e. actually $\alpha
\rightarrow \infty$. The latter means that there are no transverse
flows in the pair rest frame, saddle point in transverse velocity is
zero for fluid elements in rest, that is natural. 
Found $m_T$-scaling (for proton-$\Lambda$  pairs
$m_T=\sqrt{\left(\frac{m_p+m_{\Lambda}}{2}\right)^2+k_T^2}$) with slightly
variative multiplier $\tau$ looks very surprising.

\begin{figure}[phbt]
\center
\includegraphics[width=0.9\textwidth]{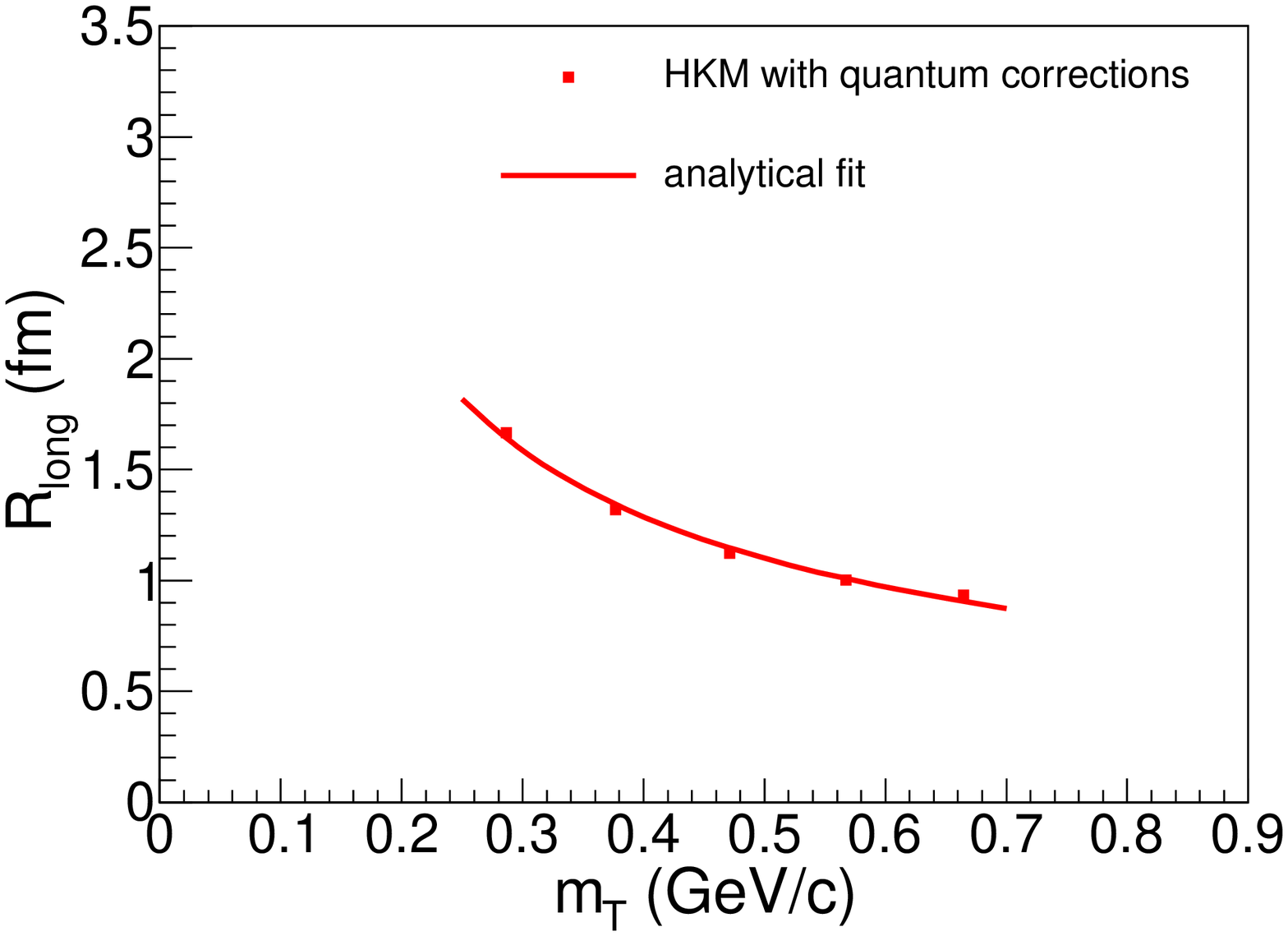}
\caption{The $m_T$ dependence of pion longitudinal femtoscopy radius $R_{l}$
calculated in the full HKM model with corrections accounting for quantum uncertainty principle
(red markers) together with the corresponding fit
(\ref{main}) (red line).
The temperature is $T=165$~MeV,
$\tau_{\mathrm{HKM}}=1.60\pm 0.04$~fm/$c$ and $\alpha=9.52\pm 0.44$.
The results are related to $\sqrt{s}=7$~TeV $pp$ collisions at the LHC, $\langle
  dN_{ch}/d\eta \rangle = 9.2$.}
\label{lhcpp9}
\end{figure}

\begin{figure}[phbt]
\center
\includegraphics[width=0.9\textwidth]{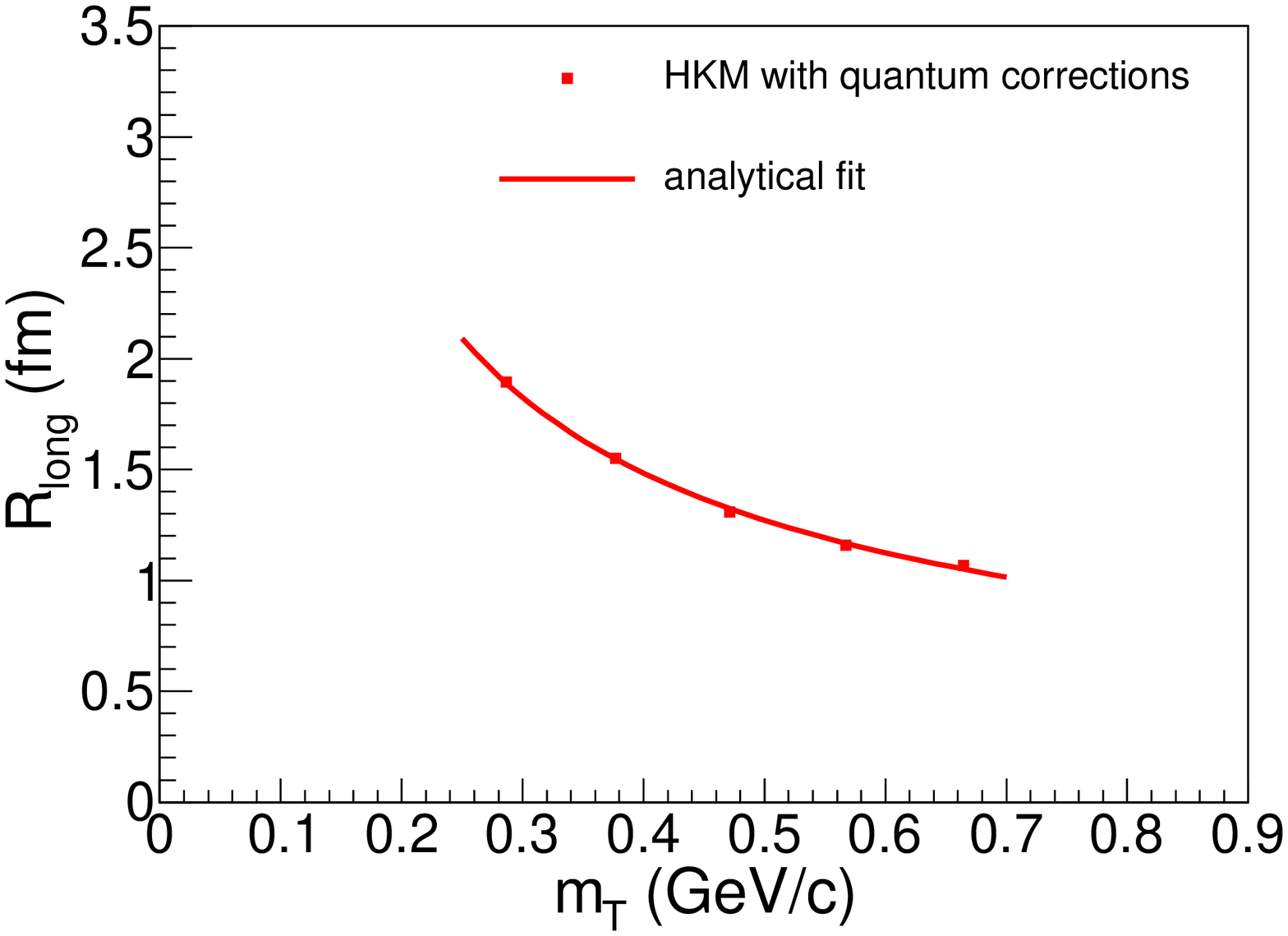}
\caption{The same as in Fig.~\ref{lhcpp9} for collision events with
$\langle dN_{ch}/d\eta \rangle = 17.9$. The values of the fit parameters  are
$\tau_{\mathrm{HKM}}=1.83\pm 0.02$~fm/$c$, $\alpha=13.9\pm 0.22$.}
\label{lhcpp17}
\end{figure}

Let us also apply the result (\ref{main}) to $p+p$ collisions at LHC energy $\sqrt{s}=7$~TeV. The theoretical results
for the pion interferometry radii in $p+p$ collisions with different particle multiplicity accounting for
uncertainty principle in small systems were presented in Ref.~\cite{PBM-Sin}.
The results without re-scattering are almost the same as in full HKM.
Therefore we assume the temperature of chemical freeze-out to be 165~MeV.
In Figs.~\ref{lhcpp9}, \ref{lhcpp17}  we fit by f-la (\ref{main})
these  results obtained in the full HKM model, including
re-scattering at the afterburner stage and corrections accounting
for quantum uncertainty principle. One can see that, as it is
expected, the values of $\tau$ in fitting function are much smaller
than in Pb+Pb collisions, namely, in the multiplicity bin with mean
charged particle multiplicity 17.9 we obtained $\tau=1.83\pm
0.02$~fm/$c$, $\alpha=13.9\pm 0.22$. At the same time for smaller
mean multiplicity 9.2 both values are smaller: $\tau=1.60\pm
0.04$~fm/c, $\alpha = 9.52\pm 0.44$.

\section{Conclusions}
The analytical formula for the longitudinal interferometry radii
that allows one to extract the time of maximal emission for the
certain hadron pairs in the case of strong transverse flows in A+A
and $p+p$ collision processes is proposed. It is compared with detail
calculations and spatiotemporal picture of the particle emission in
hydrokinetic model that already described well the large number of the
bulk observables. Within such an analysis the
factors were clarified affecting the interferometry radii and leading to 
violation of $m_T$-scaling between pions and kaons. They are: strong
transverse flows, re-scattering and resonance decays at the afterburner stage and
non-Gaussian form of the correlation function. It is found that for
compensation of the influence of the latter factor on the extracted
effective time of maximal emission for kaons, one may eliminate
restriction on the ``flow intensity'' parameter $\alpha$ for kaons
which follows from combined pion and kaon spectra fit.

As for the pion interferometry radii in $p+p$ collisions, the
analytical fits to the corresponding HKM calculations give quite
reasonable estimates for the times of the maximal emission for the
collisions with different multiplicities.

The fitting formula has been applied also to the so-called source
radii obtained from the Gaussian fit to the source function describing the
dependence of the pair production on the distance between the two
emitted particles in the rest frame of the pair. The pion-pion,
kaon-kaon, proton-proton and proton-$\Lambda$ pairs produced in
Pb+Pb collisions at the LHC were analyzed within HKM. Very
surprisingly, it was found that analytical fit with slightly
different $\tau$ for different pairs and with the other common adequate parameters --- 
no transverse flow in the pair rest frame and the same temperature taken from combined pion and
kaon spectra fit --- gives very good result and leads to the $m_T$
scaling behavior of angle-averaged source radii for all mentioned baryon and meson pairs.

We hope that the results will be especially interesting for
the experimentalists for simple fitting of the femtoscopy data and extracting
the time of maximal emission in continuously emitting
fireballs.

\section{Acknowledgment}
The authors are grateful to L.~V. Malinina for fruitful discussions and to Iu. A. Karpenko as one of the authors of HKM. The research was also carried out within the scope of the EUREA: European Ultra Relativistic Energies Agreement (European Research Group: ``Heavy ions at ultrarelativistic energies''), and is further supported by the National Academy of Sciences of Ukraine (Agreements F5-2015 and MVC-2015).

\end{document}